\documentclass[sn-mathphys]{sn-jnl}

\jyear{2021}%

\theoremstyle{thmstyleone}%
\theoremstyle{thmstyletwo}%

\theoremstyle{thmstylethree}%

\numberwithin{equation}{section}

\begin{document}

\title[Quantum Torque on a Non-Reciprocal Body out of Thermal Equilibrium]{Quantum Torque on a Non-Reciprocal Body out of Thermal Equilibrium and Induced by a Magnetic Field of Arbitrary Strength}


\author[]{\fnm{Gerard} \sur{Kennedy}}
\email{g.kennedy@soton.ac.uk}
\affil[]{\orgdiv{School of Mathematical Sciences}, \orgname{University of Southampton}, 
\orgaddress{
\city{Southampton}, \postcode{SO17 1BJ}, 
\country{United Kingdom}}}



\abstract{A stationary body that is out of thermal equilibrium with its environment, and for which the electric susceptibility is non-reciprocal, experiences a quantum torque. This arises from the spatially non-symmetric electrical response of the body to its interaction with the non-equilibrium thermal fluctuations of the electromagnetic field: the non-equilibrium nature of the thermal field fluctuations results in a net energy flow through the body, and the spatially non-symmetric nature of the electrical response of the body to its interaction with these field fluctuations causes that energy flow to be transformed into a rotational motion. We establish an exact, closed-form, analytical expression for this torque in the case that the environment is the vacuum and the material of the body is described by a damped oscillator model, where the non-reciprocal nature of the electric susceptibility is induced by an external magnetic field, as for magneto-optical media. We also generalise this expression to the context in which the body is slowly rotating. By exploring the high-temperature expansion of the torque, we are able to identify the separate contributions from the continuous spectral distribution of the non-reciprocal electric susceptibility, and from the resonance modes. In particular, we find that the torque persists in the limiting case of zero damping parameter, due to the contribution of the resonance modes. We also consider the low-temperature expansion of the torque. This work extends our previous consideration of this model to an external magnetic field of arbitrary strength, thereby including non-linear magnetic field effects. }




\maketitle

\section{Introduction}\label{sec1}

Recent years have witnessed considerable interest in non-equilibrium quantum thermodynamic phenomena, including heat transfer, torque, and non-reciprocal surface forces. For a small selection of notable papers on these topics, see  Refs. \cite{kruger2011, reid, ott2018, fogedby2018, maghrebi2019, khandekar2019, pan2019, khandekar2021, guo2020, guofan2021, gao2021, strekha2022}. 

In Ref. \cite{milton2023}, we considered the quantum torque on a body made of non-reciprocal material that is out of thermal equilibrium with its environment, an expression for which was first obtained in Ref. \cite{guofan2021}, and subsequently in Ref. \cite{strekha2022}; see also Ref. \cite{gao2021}. Broadly, such a torque arises as follows: the non-equilibrium nature of the thermal fluctuations of the electromagnetic field results in a net energy flow through the body, and the spatially non-symmetric electrical response of the body to its interaction with these field fluctuations causes that energy flow to be transformed into a rotational motion. Generally, for a single body, the existence of a quantum torque appears to require both that the system is out of thermal equilibrium and that it exhibits a broken spatial symmetry \cite{fogedby2018}.

In our previous analysis, the environment was the vacuum and the material of the body was described by a damped oscillator model, where the non-reciprocal nature of the electric susceptibility was induced by an external magnetic field, as in the case of magneto-optical media \cite{guo2020, guofan2021, gao2021}. There, we considered only contributions to first order in the magnetic field strength, which provides a good numerical approximation where the corresponding cyclotron frequency is much less than the damping parameter for the oscillator model. Here, we extend our previous analysis to consideration of an external magnetic field of arbitrary strength, thereby including non-linear magnetic field effects. 

The present paper therefore has quite a narrow focus. It is primarily concerned with investigation of the structure of the quantum torque for such a model for arbitrary external magnetic field strength, and, in particular, its spectral decomposition. Accordingly, it has a slightly more mathematical flavour than our previous analysis in Ref. \cite{milton2023}.

The remainder of this paper is structured as follows. In Section \ref{sec2}, we briefly review the damped oscillator model for the electromagnetic response of the constituents of the material of the body, and the corresponding spatially non-symmetric electric susceptibility that is induced by an external magnetic field. In Section \ref{sec3}, we consider the non-equilibrium quantum torque on a stationary such body, and establish an exact, closed-form, analytical expression for this torque that is valid for arbitrary external magnetic field strength. The corresponding torque for a slowly rotating such body is also considered. In Section \ref{sec4}, we explore the high-temperature expansion of the quantum torque for a stationary non-reciprocal body, which enables the identification of the separate contributions from the continuous spectral distribution of the non-reciprocal electric susceptibility of the body, and from the resonance modes. In particular, we find that, in the limit of vanishing damping parameter for the oscillator model, the contribution of the resonance modes results in a persistent quantum torque. The corresponding low-temperature expansion is considered in Section \ref{sec5} and brief conclusions are presented in Section \ref{sec6}. 

We use Heaviside-Lorentz electromagnetic units, and set $\hbar=c=1$, except where numerical values are presented. 

\section{Non-Reciprocal Electric Susceptibility}\label{sec2}

As in Ref. {\cite{milton2023}}, we employ the following damped oscillator model to describe the non-equilibrium displacement, $\mathbf{u}(t)$, at time $t$, of the particles that constitute the material of a body, in response to the application of a fluctuating electric field, $\mathbf{E}(t)$, and a constant external magnetic field, $\mathbf{B}$:
\begin{equation}
m\frac{d^2\mathbf{u}}{dt^2}+m\eta \frac{d \mathbf{u}}{dt}+m \omega_0^2\mathbf{u}=e\left(\mathbf{E}+\frac{d \mathbf{u}}{dt}\times \mathbf{B}\right),
\label{oscmod}
\end{equation}
where $e$ is the particle charge, $m$ is the particle mass, $\eta$ is the damping parameter, and $\omega_0$ is the free oscillation frequency. For a metal, $\omega_0=0$, and the model then reduces to that of Refs. \cite{guo2020, guofan2021}; see also Refs. \cite{seeger, zhu2014, khandekar2019, gao2021}.

This dynamical equation is easily solved in the frequency domain to yield the polarisation, $\mathbf{P}(\omega)=ne\mathbf{u}(\omega)\equiv\boldsymbol{\chi}(\omega)\mathbf{E}(\omega)$, which is here expressed in terms of the  electric susceptibility,
\begin{equation}
\boldsymbol{\chi}(\omega)=\omega_p^2
\begin{bmatrix}
\frac{\omega_0^2-\omega^2-i\omega\eta}{\left(\omega_0^2-\omega^2-i\omega\eta\right)^{\!2}-\omega^2\omega_c^2} & \frac{-i\omega\omega_c}{\left(\omega_0^2-\omega^2-i\omega\eta\right)^{\!2}-\omega^2\omega_c^2} & 0\\
 \frac{i\omega\omega_c}{\left(\omega_0^2-\omega^2-i\omega\eta\right)^{\!2}-\omega^2\omega_c^2} & \frac{\omega_0^2-\omega^2-i\omega\eta}{\left(\omega_0^2-\omega^2-i\omega\eta\right)^{\!2}-\omega^2\omega_c^2} & 0\\
0 & 0 & \frac{1}{\omega_0^2-\omega^2-i\omega\eta}
\end{bmatrix},
\label{sus}
\end{equation}
where, without loss of generality, the external magnetic field has been chosen to lie in the $z$-direction, $\omega_c\equiv \frac{e}{m}B$ is the corresponding cyclotron frequency, and $\omega_p^2\equiv \frac{ne^2}{m}$ is the square of the plasma frequency, $n$ being the particle density. 
For $B\ne 0$, this electric susceptibility is clearly non-symmetric and, moreover, non-reciprocal, that is, $\operatorname{Re}\boldsymbol{\chi}(\omega)$ is non-symmetric.

For a metal, we may set $\omega_0=0$ and use the charge and mass of the electron. For gold, the parameter values at room temperature are $\omega_p=9$ eV and $\eta =0.035$ eV \cite{hoye}. For an external magnetic field strength of 1 T, the corresponding cyclotron frequency is $\omega_c \sim 10^{-4}$ eV, which is much smaller than the damping parameter, so, in practice, keeping only terms linear in $\omega_c$ provides a good numerical approximation, one that we followed in our previous work \cite{milton2023}. Here, however, our interest is primarily in the structure of the torque, and, in particular, its spectral decomposition, rather than its magnitude. We therefore include non-linear magnetic field effects to all orders, and the expressions we derive are exact in $\omega_c$.

In the dilute limit, the mean polarisability of the body is given by 
\begin{equation}
\boldsymbol{\alpha}(\omega)=\int(d\mathbf{r})\, \boldsymbol{\chi}(\mathbf{r}; \omega),
\end{equation}
where the electric susceptibility may, in general, be position-dependent, and the integration extends over the domain of the body. Thus, for a non-reciprocal such body,\footnote{In this paper, a non-reciprocal body is defined to be one made of non-reciprocal material, that is, one for which $\operatorname{Re} \boldsymbol{\chi}(\mathbf{r};\omega)$ is non-symmetric.}  $\operatorname{Re}\boldsymbol{\alpha}(\omega)$ is, in general, non-symmetric. 

In particular, for a homogeneous such body of volume $V$ that is made of non-reciprocal material with electric susceptibility as in Eq.~(\ref{sus}) and with $\omega_0=0$, which we will henceforth assume, 
\begin{equation}
\operatorname{Re}\alpha_{xy}(\omega)=-\operatorname{Re}\alpha_{yx}(\omega)
=\frac{\omega_c\omega_p^2V}{\omega}\operatorname{Im}\left[\frac{1}{(\omega+i\eta)^2-\omega_c^2}\right],
\label{realpha1}
\end{equation}
that is, 
\begin{align}
\operatorname{Re}\alpha_{xy}(\omega)
&=-\frac{2\eta\omega_c\omega_p^2 V}{\left[(\omega+i\eta)^2-\omega_c^2\right]\left[(\omega-i\eta)^2-\omega_c^2\right]}
\nonumber\\
&=-\frac{2\eta\omega_c\omega_p^2 V}{\left[\omega^2+(\eta+i\omega_c)^2\right]\left[\omega^2+(\eta-i\omega_c)^2\right]}
\nonumber\\
&=\omega_p^2V\operatorname{Im}\left[\frac{1}{\omega^2+\xi^2}\right],
\label{realpha2}
\end{align}
where
\begin{equation}
\xi\equiv \eta+i\omega_c=\left(\eta^2+\omega_c^2\right)^{\!\frac12}e^{i\theta}, \quad \theta=\tan^{-1}\left(\frac{\omega_c}{\eta}\right).
\end{equation}
Alternatively, Eq.~(\ref{realpha1}) may be expressed in terms of partial fractions, as
\begin{equation}
\operatorname{Re}\alpha_{xy}(\omega)=-\operatorname{Re}\alpha_{yx}(\omega)=\frac{\omega_p^2V}{2\omega}\operatorname{Im}\left[\frac{1}{\omega-\omega_c+i\eta}-\frac{1}{\omega+\omega_c+i\eta}\right].
\label{realpha3}
\end{equation}

Similarly,
\begin{equation}
\operatorname{Im}\alpha_{xx}(\omega)=\operatorname{Im}\alpha_{yy}(\omega)
=-\frac{\omega_p^2V}{\omega}\operatorname{Im}\left[\frac{\omega+i\eta}{(\omega+i\eta)^2-\omega_c^2}\right],
\end{equation}
that is, 
\begin{align}
\operatorname{Im}\alpha_{xx}(\omega)
&=\frac{\omega_p^2V\eta\left(\omega^2+\eta^2+\omega_c^2\right)}{\omega\left[(\omega+i\eta)^2-\omega_c^2\right]\left[(\omega-i\eta)^2-\omega_c^2\right]}
\nonumber\\
&=\frac{\omega_p^2V\eta\left(\omega^2+\eta^2+\omega_c^2\right)}{\omega\left[\omega^2+(\eta+i\omega_c)^2\right]\left[\omega^2+(\eta-i\omega_c)^2\right]}
\nonumber\\
&=-\frac{\omega_p^2V}{2\omega\omega_c}\operatorname{Im}\left[\frac{\omega^2+\eta^2+\omega_c^2}{\omega^2+\xi^2}\right]
\nonumber\\
&=\frac{\omega_p^2V}{\omega}\operatorname{Re}\left[\frac{\xi}{\omega^2+\xi^2}\right].
\label{imalpha}
\end{align}

\section{Quantum Torque on a Non-Reciprocal Body}\label{sec3}

In Ref.~\cite{milton2023}, we used the Fluctuation-Dissipation Theorem to express the quantum torque on a stationary non-reciprocal body, with mean polarisability $\boldsymbol{\alpha}(\omega)$ in the dilute limit, as 
\begin{equation}
\tau_i^0=\int_{-\infty}^{\infty}\frac{d\omega}{2\pi} \frac{\omega^3}{6\pi}\,\epsilon_{ijk} \operatorname{Re}\alpha_{jk}(\omega) \left(\coth \frac{\beta \omega}{2}-\coth \frac{\beta'\omega}{2}\right),
\label{torquedef}
\end{equation}
where $\beta'$ is the inverse temperature of the body and $\beta$ is the inverse temperature of the environment, which is taken to be the vacuum. See also Refs. \cite{guofan2021, strekha2022}.

Thus, from Eq.~(\ref{realpha2}), the only non-zero component of the quantum torque for the damped oscillator model with $\omega_0=0$ is 
\begin{equation}
\tau_z^0=2\omega_p^2 V\int_{-\infty}^{\infty}\frac{d\omega}{2\pi} \frac{\omega^3}{6\pi}\operatorname{Im}\left[\frac{1}{\omega^2+\xi^2}\right] \left(\coth \frac{\beta \omega}{2}-\coth \frac{\beta'\omega}{2}\right),
\label{torqueexp1}
\end{equation}
which becomes
\begin{align}
\tau_z^0&=8\omega_p^2 V\int_{0}^{\infty}\frac{d\omega}{2\pi} \frac{\omega^3}{6\pi}\operatorname{Im}\left[\frac{1}{\omega^2+\xi^2}\right] \left(\frac{1}{e^{\beta\omega}-1}-\frac{1}{e^{\beta'\omega}-1}\right)
\nonumber\\
&=-\frac{2\omega_p^2V}{3\pi^2}\operatorname{Im}\left[\xi^2\int_0^{\infty}d\omega\,\frac{\omega}{\omega^2+\xi^2}\left(\frac{1}{e^{\beta\omega}-1}-\frac{1}{e^{\beta'\omega}-1}\right)\right],
\end{align}
that is, 
\begin{align}
\tau_z^0&=\frac{\omega_p^2V}{3\pi^2}\Biggl\{\pi\omega_c\left(\frac{1}{\beta}-\frac{1}{\beta'}\right)-2\eta\omega_c\log\left(\frac{\beta}{\beta'}\right)
\nonumber\\
&\qquad\quad
+\operatorname{Im}\left[\xi^2\left(\psi\left(\frac{\beta\xi}{2\pi}\right)-\psi\left(\frac{\beta' \xi}{2\pi}\right)\right)\right]\Biggr\},
\label{torqueexp}
\end{align}
where we have employed the integral representation of the digamma function \cite{whitwat, erdelyi, gradshteyn, askey}, 
\begin{equation}
\psi(s)=\log s -\frac{1}{2s}-2\int_0^{\infty}dt\,\frac{t}{(t^2+s^2)\left(e^{2\pi t}-1\right)}, \qquad \operatorname{Re}s >0.
\end{equation}

Keeping only terms to first order in $\omega_c$, it is easily verified that Eq.~(\ref{torqueexp}) agrees with Eq.~(4.15) of our earlier work \cite{milton2023}.

For a sense of scale, note that for a gold nanosphere of radius 100 nm, with the parameter values given in Section \ref{sec2}, in an external magnetic field of strength 1~T, the magnitude of the coefficient of the $\log$ term in Eq.~(\ref{torqueexp}) is  $1.6 \times 10^{-24}$ Nm. 

The extension to a slowly rotating body was also considered in Ref.~\cite{milton2023}; see also Ref. \cite{guofan2021}. There, the $z$-component of the quantum torque on a non-reciprocal body undergoing a non-relativistic rotation about the $z$-axis with angular frequency $\Omega$, and with mean polarisability $\boldsymbol{\alpha}(\omega)$ in the dilute limit, was expressed as
\begin{align}
\tau_z&=\int_{-\infty}^{\infty}\frac{d\omega}{2\pi}\frac{\omega_+^3}{6\pi}\left[\operatorname{Im}\left(\alpha_{xx}+\alpha_{yy}\right)(\omega)+\operatorname{Re}\left(\alpha_{xy}-\alpha_{yx}\right)(\omega)\right]
\nonumber\\
&\qquad\qquad\qquad \times\left(\coth \frac{\beta\omega_+}{2}-\coth\frac{\beta'\omega}{2}\right), 
\end{align}
where $\omega_+\equiv \omega+\Omega$. Keeping only terms to first order in $\Omega$, we may write 
\begin{equation}
\tau_z=\tau_z^0+\Omega\,\tau_z^1,
\label{torquerotexp}
\end{equation}
where $\tau_z^0$ is given by Eq.~(\ref{torquedef}) and 
\begin{equation}
\tau_z^1=\left(3+\beta\frac{\partial}{\partial \beta}\right) \hat\tau_z^1
\end{equation}
with
\begin{equation}
\hat\tau_z^1\equiv  \int_{-\infty}^{\infty}\frac{d\omega}{2\pi}\frac{\omega^2}{6\pi}
\operatorname{Im}\left(\alpha_{xx}+\alpha_{yy}\right)(\omega)
\left(\coth \frac{\beta\omega}{2}-\coth\frac{\beta'\omega}{2}\right).
\end{equation}

Thus, from Eq.~(\ref{imalpha}) for the damped oscillator model with $\omega_0=0$, 
\begin{align}
\hat\tau_z^1&=2\omega_p^2 V \int_{-\infty}^{\infty}\frac{d\omega}{2\pi} \frac{\omega}{6\pi}\operatorname{Re}\left[\frac{\xi}{\omega^2+\xi^2}\right] \left(\coth \frac{\beta \omega}{2}-\coth \frac{\beta'\omega}{2}\right)
\nonumber\\
&=8\omega_p^2 V \int_{0}^{\infty}\frac{d\omega}{2\pi} \frac{\omega}{6\pi}\operatorname{Re}\left[\frac{\xi}{\omega^2+\xi^2}\right] \left(\frac{1}{e^{\beta\omega}-1}-\frac{1}{e^{\beta'\omega}-1}\right)
\nonumber\\
&=\frac{2\omega_p^2V}{3\pi^2}\operatorname{Re}\left[\xi\int_0^{\infty}d\omega\,\frac{\omega}{\omega^2+\xi^2}\left(\frac{1}{e^{\beta\omega}-1}-\frac{1}{e^{\beta'\omega}-1}\right)\right],
\end{align}
that is, 
\begin{align}
\hat\tau_z^1&=\frac{\omega_p^2V}{3\pi^2}\Biggl\{-\pi\left(\frac{1}{\beta}-\frac{1}{\beta'}\right)+\eta\log\left(\frac{\beta}{\beta'}\right)
-\operatorname{Re}\left[\xi\left(\psi\left(\frac{\beta \xi}{2\pi}\right)-\psi\left(\frac{\beta' \xi}{2\pi}\right)\right)\right]\Biggr\},
\end{align}
whence,
\begin{align}
\tau_z^1&=\frac{\omega_p^2V}{3\pi^2}\Biggl\{-\pi\left(\frac{2}{\beta}-\frac{3}{\beta'}\right)+3\eta\log\left(\frac{\beta}{\beta'}\right)+\eta
\nonumber\\
&\qquad\quad-3\operatorname{Re}\left[\xi\left(\psi\left(\frac{\beta\xi}{2\pi}\right)-\psi\left(\frac{\beta' \xi}{2\pi}\right)\right)\right]
-\frac{\beta}{2\pi}\operatorname{Re}\left[\xi^2\, \psi'\left(\frac{\beta \xi}{2\pi}\right)\right]\Biggr\}.
\label{torque1exp}
\end{align}

Keeping only terms to first order in $\omega_c$, it is easily verified that Eq.~(\ref{torque1exp}) agrees with Eq.~(5.10) of our earlier work \cite{milton2023}. 

The quantum torque on a rotating non-reciprocal body, made of material described by the damped oscillator model with $\omega_0=0$, is therefore given, to first order in $\Omega$,  by Eq.~(\ref{torquerotexp}), together with Eqs.~(\ref{torqueexp}) and (\ref{torque1exp}).

\section{High-Temperature Expansion}\label{sec4}

It is instructive to consider the high-temperature expansion of the quantum torque. We do so here only for a stationary body; the same approach could also be applied to a slowly rotating body. 

For high temperatures, $\beta, \beta' \to 0^+$, use of the power series representation \cite{erdelyi, gradshteyn, askey}, 
\begin{equation}
\psi(s)=-\frac{1}{s}-\gamma+\sum_{n=2}^{\infty}(-1)^n\zeta(n) \,s^{n-1}, \qquad \vert s\vert <1,
\label{psismalls}
\end{equation}
in Eq.~(\ref{torqueexp}) yields
\begin{align}
\tau_z^0&=\frac{\omega_p^2V}{3\pi^2}\Biggl\{-\pi\omega_c\left(\frac{1}{\beta}-\frac{1}{\beta'}\right)-2\eta\omega_c\log\left(\frac{\beta}{\beta'}\right)
\nonumber\\
&\qquad\quad+\sum_{n=2}^{\infty}\frac{(-1)^n\zeta(n)}{(2\pi)^{n-1}}\,\xi_{n+1}\left(\beta^{n-1}-\beta'^{n-1}\right)\Biggr\},
\label{hightempexp}
\end{align}
where
\begin{equation}
\xi_k\equiv \operatorname{Im}\left[\xi^{k}\right]=\left(\eta^2+\omega_c^2\right)^{\!\frac{k}{2}}\sin{\left(k\theta\right)}, \qquad k \in \mathbb{Z}.
\end{equation}

It is easily seen that $\xi_k$ satisfies the reflection formula
\begin{equation}
\xi_{-k}=-\left(\eta^2+\omega_c^2\right)^{-k}\xi_{k}, \qquad k\in \mathbb{Z},
\label{xireflect}
\end{equation}
and the recurrence relation
\begin{equation}
\xi_k=2\eta\,\xi_{k-1}-\left(\eta^2+\omega_c^2 \right)\xi_{k-2}, \qquad k\in \mathbb{Z}.
\label{xirecurr}
\end{equation}
The latter may be used to recursively generate $\xi_k$. For example, the first few $\xi_k$ for non-negative $k$, generated in this way, are as follows:
\begin{align}
\xi_0 &= 0,\\
\xi_1 &= \omega_c,\\
\xi_2 &= 2\eta\omega_c,\\
\xi_3 &= 3\eta^2\omega_c-\omega_c^3,\\
\xi_4 &= 4\eta^3\omega_c-4\eta\omega_c^3,\\
\xi_5 &= 5\eta^4\omega_c-10\eta^2\omega_c^3+\omega_c^5,\\
\xi_6 &= 6\eta^5\omega_c-20\eta^3\omega_c^3+6\eta\omega_c^5.
\end{align}

Alternatively, these expressions may be established directly by using a suitable multiple-angle formula, as in the following, for $k \ge 1$:
\begin{align}
\xi_k&=\left(\eta^2+\omega_c^2\right)^{\!\frac{k}{2}}\sin\theta \,U_{k-1}(\cos \theta)
\nonumber\\
&=\omega_c\sum_{r=0}^{\left\lfloor \frac{k-1}{2}\right\rfloor} (-1)^r \binom{k-1-r}{r} (2\eta)^{k-1-2r}\left(\eta^2+\omega_c^2\right)^r
\nonumber\\
&=\sum_{m=0}^{\left\lfloor \frac{k-1}{2}\right\rfloor} (-1)^m \binom{k}{2m+1}\,\eta^{k-1-2m}\,\omega_c^{2m+1},
\label{cheby}
\end{align}
where $U_n$ denotes the Chebyshev polynomial of the second kind of order $n$. 

It follows immediately from Eq.~(\ref{xirecurr}), or from Eqs.~(\ref{cheby}) and (\ref{xireflect}), that, in the limit of vanishing damping parameter, the behaviour of $\xi_k$ differs, depending on whether the index, $k\in \mathbb{Z}$, is even or odd:
\begin{equation}
\lim_{\eta\to 0^+} \xi_{k}=
\begin{cases}
0, &\text{for $k$ even},\\
(-1)^{\frac{k-1}{2}}\omega_c^k, &\text{for $k$ odd}.
\label{xilimit}
\end{cases}
\end{equation}
As a consequence, only the odd $\beta, \beta'$ powers in Eq.~(\ref{hightempexp}) survive in this limit, resulting in 
\begin{align} 
\lim_{\eta\to 0^+}\tau_z^0&= \frac{\omega_p^2V}{3\pi^2}\sum_{m=0}^{\infty}\frac{(-1)^m\zeta(2m)}{(2\pi)^{2m-1}}\,\omega_c^{2m+1}\left(\beta^{2m-1}-\beta'^{2m-1}\right)
\nonumber\\
&=-\frac{\omega_p^2V}{3\pi}\sum_{m=0}^{\infty}\frac{B_{2m}}{(2m)!}\,\omega_c^{2m+1}\left(\beta^{2m-1}-\beta'^{2m-1}\right).
\label{taulimit}
\end{align}

It may be slightly surprising to note that $\lim_{\eta\to 0^+}\tau_z^0\ne 0$, since it is clear from Eq.~(\ref{realpha1}) that if $\eta=0$ then $\operatorname{Re}\alpha_{xy}(\omega)=0$, and it then follows from Eq.~(\ref{torquedef}) that $\tau_z^0=0$. That is, $\tau_z^0$ is a discontinuous function of $\eta$ as $\eta \to 0^+$.

In fact, this behaviour is due to the existence of resonance modes at $\omega=\pm\omega_c -i\eta$ in the complex-$\omega$ plane, corresponding to poles of the integrand in Eq.~(\ref{torqueexp1}). In the limit $\eta\to 0^+$, these poles approach the real line from below, and the integral in Eq.~(\ref{torqueexp1}) then becomes formally divergent. However, it may be assigned a finite value by invoking the Sokhotski-Plemelj Theorem, which may be written in the form
\begin{equation}
\lim_{\eta\to 0^+}\frac{1}{x\pm i\eta}= {\mathcal P}\left(\frac{1}{x}\right)\mp i\pi\delta(x),
\label{SPT}
\end{equation}
where $\mathcal P$ denotes the corresponding Cauchy Principal Value (CPV) integral. See, for example, Ref. \cite{galapon2016}, where it is shown that the CPV integral may, equivalently, be expressed as an Analytic Principal Value integral. 

Indeed, using the partial fraction representation for $\operatorname{Re}\alpha_{xy}(\omega)$ given in Eq.~(\ref{realpha3}), Eq.~(\ref{torquedef}) may be alternatively written as 
\begin{equation}
\tau_z^0=\operatorname{Im}\int_{-\infty}^{\infty}d\omega\left[\frac{1}{\omega-\omega_c+i\eta}-\frac{1}{\omega+\omega_c+i\eta}\right]f(\omega),
\label{torquedefpf}
\end{equation}
where
\begin{equation}
f(\omega)\equiv \frac{\omega_p^2V}{12\pi^2}\, \omega^2 \left(\coth \frac{\beta\omega}{2}-\coth \frac{\beta'\omega}{2}\right).
\end{equation}
Echoing the decomposition in Eq.~(\ref{SPT}), Eq.~(\ref{torquedefpf}) has the limit
\begin{equation}
\lim_{\eta\to 0^+}\tau_z^0= \lim_{\eta\to 0^+}\tau_z^{0\mathcal P}+\lim_{\eta\to 0^+}\tau_z^{0\mathcal R},
\end{equation}
where the CPV contribution, indicated by $\mathcal P$, and the resonance contribution, indicated by $\mathcal R$, are respectively given by 
\begin{equation}
\lim_{\eta\to 0^+}\tau_z^{0\mathcal P}=0,
\label{tauplimit}
\end{equation}
since the integrand in Eq.~(\ref{torquedefpf}) is real in this limit, and 
\begin{align}
\lim_{\eta\to 0^+}\tau_z^{0\mathcal R}&=-\frac{\omega_c^2\omega_p^2V}{6\pi}\left(\coth \frac{\beta\omega_c}{2}-\coth \frac{\beta'\omega_c}{2}\right)
\nonumber\\
&=-\frac{\omega_c^2\omega_p^2V}{3\pi}\left(\frac{1}{e^{\beta\omega_c}-1}-\frac{1}{e^{\beta'\omega_c}-1}\right).
\label{taureslimit}
\end{align}
It is easily verified, using the exponential generating function of the Bernoulli numbers, 
\begin{equation}
\frac{x}{e^x-1}=\sum_{k=0}^{\infty}\frac{B_k}{k!}x^k, 
\label{Bernoulli}
\end{equation}
that Eq.~(\ref{taureslimit}) is a more compact equivalent of Eq.~(\ref{taulimit}). Thus, the fact that $\lim_{\eta\to 0^+} \tau_z^0 \ne 0$ is due entirely to the contribution of the resonance modes.

The persistence of the quantum torque in the limit of vanishing damping found here is reminiscent of similar behaviour found in our earlier work \cite{milton2020} for a charged particle passing a conducting plate with permittivity governed by the Drude model. There, the transverse magnetic component of the classical electrodynamic friction on the particle was found to persist in the limit of vanishing damping (resistivity) parameter, again due to the presence, in this limit, of a singularity in the integrand for the friction. 

\begin{figure}[H]
 \centering
\includegraphics[width=0.85\linewidth]{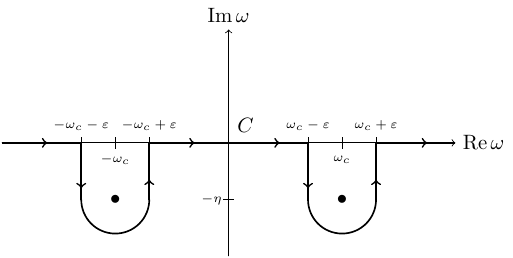}
\caption{The contour $C$ used for the evaluation of $\tau_z^0$. The corresponding CPV integral over $C$ excludes the semi-circles around the poles at $\omega=\pm \omega_c-i\eta$, and is evaluated in the limit $\varepsilon \to 0^+$.}
\label{contour}
\end{figure}

Let us now generalise the above to finite $\eta$. Using the residue theorem to replace the integral over the real line in Eq.~(\ref{torquedefpf}) by that over the contour $C$ displayed in  Figure~\ref{contour}, we may write
\begin{align}
\tau_z^0&=\operatorname{Im}\biggl\{\int_C d\omega\left[\frac{1}{\omega-\omega_c+i\eta}-\frac{1}{\omega+\omega_c+i\eta}\right] f(\omega)
\nonumber\\
&\qquad\quad-2\pi i \left[f(\omega_c-i\eta)-f(-\omega_c-i\eta)\right]\biggr\}
\nonumber\\
&=\operatorname{Im}\biggl\{{\mathcal P}\int_{C}d\omega\left[\frac{1}{\omega-\omega_c+i\eta}-\frac{1}{\omega+\omega_c+i\eta}\right] f(\omega)
\nonumber\\
&\qquad\quad-\pi i\left[f(\omega_c-i\eta)-f(-\omega_c-i\eta)\right]\biggr\}
\nonumber\\
&={\mathcal P}\int_{-\infty}^{\infty}d\omega\,\operatorname{Im}\left[\frac{1}{\omega-\omega_c+i\eta}-\frac{1}{\omega+\omega_c+i\eta}\right]f(\omega)
\nonumber\\
&\qquad\quad-\pi\left[f(\omega_c-i\eta)+f(\omega_c+i\eta)\right],
\end{align}
where the CPV integral over the contour $C$ excludes the semi-circles around the poles at $\omega=\pm\omega_c -i\eta$, and is evaluated in the limit $\varepsilon \to 0^+$, where it is equal to the corresponding integral over the real line, excluding the points $\omega=\pm\omega_c$. 

Thus, we may write
\begin{equation}
\tau_z^0=\tau_z^{0\mathcal P}+\tau_z^{0\mathcal R}, 
\label{tausum}
\end{equation}
where
\begin{align}
\tau_z^{0\mathcal P}&\equiv\frac{\omega_p^2V}{12\pi^2}\,{\mathcal P}\int_{-\infty}^{\infty}\!d\omega\,\omega^2\operatorname{Im}\!\left[\frac{1}{\omega-\omega_c+i\eta}-\frac{1}{\omega+\omega_c+i\eta}\right]
\nonumber\\
&\qquad\qquad\times\left(\coth \frac{\beta\omega}{2}-\coth\frac{\beta'\omega}{2}\right)
\label{taupexp}
\end{align}
and
\begin{align}
\tau_z^{0\mathcal R}&\equiv -\frac{\omega_p^2V}{6\pi}\operatorname{Re}\left[\left(\omega_c+i\eta\right)^2\left(\coth \frac{\beta\left(\omega_c+i\eta\right)}{2}-\coth\frac{\beta'\left(\omega_c+i\eta\right)}{2}\right)\right]
\nonumber\\
&=-\frac{\omega_p^2V}{3\pi}\operatorname{Re}\left[\left(\omega_c+i\eta\right)^2\left(\frac{1}{e^{\beta\left(\omega_c+i\eta\right)}-1}-\frac{1}{e^{\beta'\left(\omega_c+i\eta\right)}-1}\right)\right].
\label{tauresexp}
\end{align}

Again using Eq.~(\ref{Bernoulli}), it is easily verified that Eq.~(\ref{tauresexp}) may be written as
\begin{equation}
\tau_z^{0\mathcal R}=-\frac{\omega_p^2V}{3\pi}\sum_{m=0}^{\infty}\frac{(-1)^mB_{2m}}{(2m)!}\,\xi_{2m+1}\left(\beta^{2m-1}-\beta'^{2m-1}\right),
\end{equation}
which agrees with the odd $\beta, \beta'$ powers in Eq.~(\ref{hightempexp}). Also, it is immediate that Eq.~(\ref{tauresexp}) agrees with Eq.~(\ref{taureslimit}) in the limit $\eta \to 0^+$. 

To evaluate Eq.~(\ref{taupexp}), it suffices to again express the integrand in the form seen in Eq.~(\ref{torqueexp1}), but accommodate the CPV by using the high-frequency expansion of $\operatorname{Im}\left[\frac{1}{\omega^2+\xi^2}\right]$ away from the poles at $\omega=\pm\omega_c-i\eta$. This results in
\begin{align}
\tau_z^{0\mathcal P}&=\frac{2\omega_p^2V}{3\pi^2}\sum_{k=0}^{\infty}(-1)^k\xi_{2k}\int_0^{\infty}d\omega\,\omega^{1-2k}\left(\frac{1}{e^{\beta\omega}-1}-\frac{1}{e^{\beta'\omega}-1}\right)
\nonumber\\
&=\frac{2\omega_p^2V}{3\pi^2}\sum_{k=0}^{\infty}(-1)^k \xi_{2k}\Gamma(2-2k)\zeta(2-2k) \left(\beta^{2k-2}-\beta'^{2k-2}\right)
\nonumber\\
&=\frac{\omega_p^2V}{3\pi^2}\Biggl\{-2\eta\omega_c\log\left(\frac{\beta}{\beta'}\right)-\sum_{m=1}^{\infty}
\frac{\zeta(2m+1)}{(2\pi)^{2m}}\,\xi_{2m+2}\left(\beta^{2m}-\beta'^{2m}\right)\Biggr\},
\label{taupexp2}
\end{align}
where we have used the functional equation
\begin{equation}
\zeta(s)=2 (2\pi)^{s-1}\sin\left(\frac{\pi s}{2}\right)\Gamma(1-s)\zeta(1-s)
\end{equation}
to evaluate the formally divergent integrals by analytic continuation and have interpreted the $k=1$ term as the limit, 
\begin{equation}
\lim_{k\to 1}\Gamma(2-2k)\zeta(2-2k) \left(\beta^{2k-2}-\beta'^{2k-2}\right)=\frac12 \log\left(\frac{\beta}{\beta'}\right). 
\end{equation}
See Ref. \cite{galapon2017} for interpretation of the finite part of the divergent integrals that may result from term-by-term integration, and how missing terms in such integration arise from singularities of the integrand. As expected, Eq.~(\ref{taupexp2}) agrees with the even $\beta, \beta'$ powers, and logarithms, in Eq.~(\ref{hightempexp}), and, from Eq.~(\ref{xilimit}), with Eq.~(\ref{tauplimit}) in the limit $\eta\to 0^+$. 

So, we may conclude that the continuous spectral distribution of $\operatorname{Re}\alpha_{xy}(\omega)$ in Eq.~(\ref{torquedef}) generates the even $\beta, \beta'$ powers, and logarithms, in the high-temperature expansion of the quantum torque, Eq.~(\ref{hightempexp}), while the discrete spectral distribution of $\operatorname{Re}\alpha_{xy}(\omega)$, arising from the resonance modes, generates the odd $\beta, \beta'$ powers. 

However, it should be noted that, like the power series representation, Eq.~(\ref{psismalls}), Eq.~(\ref{hightempexp}) is an exact expression for the quantum torque, simply one that has been written in a form that is suitable for high temperature expansion. Thus, the evenness and oddness of the two types of contribution above under change of sign of the inverse temperatures is a property that enables closed-form expressions for these two types of contribution to be easily identified. 

We may therefore write
\begin{equation}
\tau_z^{0\mathcal P}=\frac{\tau_z^0\left(\beta, \beta'\right)+\tau_z^0\left(-\beta,-\beta'\right)}{2}
\end{equation}
and
\begin{equation}
\tau_z^{0\mathcal R}=\frac{\tau_z^0\left(\beta, \beta'\right)-\tau_z^0\left(-\beta,-\beta'\right)}{2},
\end{equation}
which, from Eq.~(\ref{torqueexp}), generate the corresponding closed-form expressions
\begin{align}
\tau_z^{0\mathcal P}&=\frac{\omega_p^2V}{6\pi^2}\Biggl\{-4\eta\omega_c\log\left(\frac{\beta}{\beta'}\right)
\nonumber\\
&\quad
+\operatorname{Im}\left[\xi^2\left(\psi\left(\frac{\beta\xi}{2\pi}\right)+\psi\left(-\frac{\beta\xi}{2\pi}\right)
-\psi\left(\frac{\beta' \xi}{2\pi}\right)-\psi\left(-\frac{\beta' \xi}{2\pi}\right)
\right)\right]\Biggr\}
\end{align}
and
\begin{align}
\tau_z^{0\mathcal R}&=\frac{\omega_p^2V}{6\pi^2}\Biggl\{2\pi\omega_c\left(\frac{1}{\beta}-\frac{1}{\beta'}\right)
\nonumber\\
&\quad
+\operatorname{Im}\left[\xi^2\left(\psi\left(\frac{\beta\xi}{2\pi}\right)-\psi\left(-\frac{\beta\xi}{2\pi}\right)
-\psi\left(\frac{\beta' \xi}{2\pi}\right)+\psi\left(-\frac{\beta' \xi}{2\pi}\right)
\right)\right]\Biggr\}.
\label{tauresexp2}
\end{align}
It is easily verified from the reflection formula \cite{erdelyi}
\begin{equation}
\psi(-s)=\psi(s)+\pi\cot (\pi s) +\frac1s
\end{equation}
that Eq.~(\ref{tauresexp2}) agrees with the more compact equivalent of Eq.~(\ref{tauresexp}).

Evenness or oddness under change of sign of the inverse temperatures is suggestive of similar behaviour under time reversal, since the corresponding thermal Green functions are periodic in imaginary time, with period equal to the relevant inverse temperature. Indeed, the model in Eq.~(\ref{oscmod}) is invariant under time reversal if we also reverse the signs of both $\eta$ and $\omega_c$, the latter corresponding to change in the sign of $\mathbf{B}$. It is easily seen from Eq.~(\ref{cheby}) and Eq.~(\ref{xireflect}) that, for $k\in \mathbb{Z}$, 
\begin{equation}
\xi_k(-\eta)=(-1)^{k-1}\xi_k(\eta) \quad\text{and}\quad \xi_k(-\omega_c)=-\xi_k(\omega_c), 
\end{equation}
whence
\begin{equation}
\xi_k(-\eta, -\omega_c)=(-1)^k \xi_k(\eta, \omega_c). 
\end{equation}
Thus, alternatively, the two types of contribution to the quantum torque may be extracted by writing
\begin{equation}
\tau_z^{0\mathcal{P}}=\frac{\tau_z^0(\eta, \omega_c)+\tau_z^0(-\eta, -\omega_c)}{2}
\end{equation}
and
\begin{equation}
\tau_z^{0\mathcal{R}}=\frac{\tau_z^0(\eta, \omega_c)-\tau_z^0(-\eta, -\omega_c)}{2}.
\end{equation}

\section{Low-Temperature Expansion}\label{sec5}

The low-temperature expansion of the quantum torque is less interesting, but is still worth considering. Again, for illustration, we do so here only for a stationary body. 

For low temperatures, $\beta, \beta' \to \infty$, use of the asymptotic series representation \cite{erdelyi, askey}, 
\begin{equation}
\psi(s) \sim \log s -\frac{1}{2s}-\sum_{n=1}^{\infty}\frac{B_{2n}}{2n s^{2n}}, \qquad \lvert\arg s\rvert <\pi, s\to\infty, 
\end{equation}
in Eq.~(\ref{torqueexp}) yields
\begin{equation}
\tau_z^0 \sim -\frac{\omega_p^2V}{3\pi^2}\sum_{n=2}^{\infty}\frac{(2\pi)^{2n}B_{2n}}{2n}\,\xi_{2-2n}\left(\frac{1}{\beta^{2n}}-\frac{1}{\beta'^{2n}}\right).
\label{lowtempexp}
\end{equation}

On the other hand, use of the low-frequency expansion of $\operatorname{Im}\left[\frac{1}{\omega^2+\xi^2}\right]$ in the integrand of Eq.~(\ref{torqueexp1}) results in
\begin{align}
\tau_z^{0\mathcal P}&= \frac{2\omega_p^2V}{3\pi^2}\sum_{k=0}^{\infty}(-1)^k\xi_{-2-2k}\int_0^{\infty}d\omega\,\omega^{3+2k}\left(\frac{1}{e^{\beta\omega}-1}-\frac{1}{e^{\beta'\omega}-1}\right)
\nonumber\\
&=\frac{2\omega_p^2V}{3\pi^2}\sum_{k=0}^{\infty}(-1)^k \xi_{-2-2k}\,\Gamma(4+2k)\, \zeta(4+2k) \left(\frac{1}{\beta^{4+2k}}-\frac{1}{\beta'^{4+2k}}\right)
\nonumber\\
&=-\frac{\omega_p^2V}{3\pi^2}\sum_{n=2}^{\infty}\frac{(2\pi)^{2n}B_{2n}}{2n}\,\xi_{2-2n}\left(\frac{1}{\beta^{2n}}-\frac{1}{\beta'^{2n}}\right),
\label{taulowfreq}
\end{align}
in agreement with the asymptotic expression, Eq.~(\ref{lowtempexp}).

So, for low temperatures, $\tau_z^0$ may again be decomposed as in Eq.~(\ref{tausum}), with the CPV contribution as in Eq.~(\ref{taulowfreq}) and the resonance contribution as in Eq.~(\ref{tauresexp}), but, in this case, the resonance contribution is exponentially damped (that is, the resonance modes are not excited at low temperatures), and so it does not feature in the asymptotic representation, Eq.~(\ref{lowtempexp}).

\section{Conclusion}\label{sec6}

In this paper, we have considered the quantum torque experienced by a non-reciprocal body that is out of thermal equilibrium with its environment, where the material of the body is described by a damped oscillator model and where the non-reciprocal nature of the electric susceptibility is induced by an external magnetic field of arbitrary strength, thereby extending our previous consideration of this model \cite{milton2023} to include non-linear magnetic field effects. 

We have established an exact, closed-form, analytical expression for the quantum torque on a stationary such body, and have generalised this expression to the context in which the body is slowly rotating. 

By exploring the high-temperature expansion of the quantum torque on a stationary non-reciprocal body, we have been able to identify the separate contributions to this torque arising from the continuous spectral distribution of the electric susceptibility, and from the resonance modes. These two types of contribution have been found to differ in parity: that from the continuous spectral distribution is even under combined reflection of the inverse temperatures of the body and of its environment, while that from the resonance modes is odd under such reflection. 

In particular, we have demonstrated that, while the contribution to the quantum torque due to the continuous spectral distribution vanishes in the limit of vanishing damping parameter, that due to the resonance modes does not, resulting in a persistent quantum torque in this limit. The quantum torque is, therefore, a discontinuous function of the damping parameter in this limit.

We have also considered the low-temperature expansion of the quantum torque on a stationary non-reciprocal body, where the two types of contribution again display the parity characteristics described above, but where that due to the resonance modes is exponentially damped. 

There are, of course, limitations to our analysis. In particular, we have considered the mean polarisability of the non-reciprocal body only in the dilute limit, and have assumed that the temperatures of the body and of its environment do not change with time, so that the non-equilibrium nature of the configuration is maintained. These limitations are discussed in our previous work \cite{milton2023}, and are not explored further here.

In addition, for simplicity, we have restricted attention only to the case in which the environment is the vacuum. It would be interesting to extend this work to encompass more general environments, such as a dielectric medium, to explore related phenomena, such as angular momentum transfer, and to consider more realistic physical models for nonreciprocal bodies, including magneto-optical media.

\backmatter

\bmhead{Acknowledgments}

This work was supported in part by a grant from the U.S. National Science Foundation, No. PHY-1748958, which enabled the author to participate in the 2022 workshop {\it Emerging Regimes and Implications of Quantum and Thermal Fluctuational Electrodynamics}, held at the Kavli Institute for Theoretical Physics (KITP). This workshop stimulated the research described here, and the author thanks KITP and the organisers of the workshop for their kind hospitality. The author is indebted to Kimball Milton, from whom he learned almost all that he knows regarding quantum torque and non-reciprocal media, and thanks Kimball Milton and Stephen Fulling for their helpful comments on an earlier version of this paper. For the purpose of open access, the author has applied a CC BY public copyright licence to any Author Accepted Manuscript version arising from this submission.







\end{document}